\documentstyle[aps,eqsecnum,12pt]{revtex}

\title{
Equation of motion for relativistic compact binaries
\\ with the strong field point particle limit :\\
Formulation,
the first post-Newtonian and multipole terms
}

\author{Yousuke Itoh \footnote{yousuke@astr.tohoku.ac.jp}, 
Toshifumi Futamase\footnote{tof@astr.tohoku.ac.jp}\\
Astronomical Institute, Graduate School of 
Science, Tohoku University \\
Sendai 980-8578, Japan \\ 
and \\
Hideki Asada\footnote{asada@phys.hirosaki-u.ac.jp}, \\
Faculty of Science and Technology, 
Hirosaki University, \\
Hirosaki 036-8561, Japan
}

\begin{document}
\maketitle

\begin{abstract} 

We derive the equation of motion for the relativistic compact binaries 
in the post-Newtonian approximation taking  explicitly 
their strong internal gravity into account. For this purpose we 
adopt the method of the point particle limit where the equation 
of motion  is expressed in terms of the surface integrals.  
We examine carefully the behavior of the surface integrals 
in the derivation. 
As a result, we obtain the Einstein-Infeld-Hoffman equation of
 motion at the first 
post-Newtonian (1PN) order, and a part of the 2PN order 
which depends on the quadrupole moments and the spins of component stars. 
Hence, it is found that the equation of motion in the post-Newtonian 
approximation is valid for the compact binaries by a suitable 
definition of the mass, spin and quadrupole moment. 
\end{abstract}

\begin{flushleft}
PACS Number(s): 04.25.Nx, 
\end{flushleft}

\newcommand{\pa}{\partial}

\section{Introduction} 
The motion of self gravitating systems and the associated 
gravitational waves have been among the  major 
research interest 
in general relativity. 
Furthermore, understanding systems with strong internal gravity such as 
neutron star binaries are important 
also in astrophysics, since they are thought as a possible candidate for
 gamma
 ray bursts \cite{Paczynski} as well as the most likely sources of 
gravitational waves. 
Laser interferometric detectors are under construction 
for detecting gravitational waves 
(LIGO \cite{Abramovici}, VIRGO \cite{Bradaschia}, 
TAMA \cite{Kuroda} and GEO600 \cite{Hough}). 
It is expected that these will become new eyes looking into the
universe and bring great impacts on astrophysics. 
For this to become realistic,  reliable comparisons  between
theoretical predictions 
and the observations are definitely necessary. 
Namely we have to  prepare sufficiently accurate 
theoretical templates for waveforms for likely sources 
including inspiralling neutron star binaries. 
This is possible only if we know their precise equation 
of motion \cite{BDIWW,Blanchet,MSSTT,AF}. 
Since the Einstein equation is highly nonlinear system as is well-known, 
the use of some kind of approximation is unavoidable to 
describe a dynamical system such as a neutron stars binary. 

Among various approximation schemes, the post-Newtonian approximation is 
most popular  for treating nonlinear gravitational interactions 
\cite{CE}, and the equation of motion has been derived successfully 
up to the 2.5PN order \cite{DD81a,DD81b,Damour,Kopejkin,BFP}. 
At higher orders, however, an expansion in power series seems to 
break down (e. g. \cite{FS83}). 
Hence, it is very interesting to see what form the equation of motion 
takes at higher orders.  
In recent, 3PN, 3.5PN and higher orders are being tackled \cite{JS,GII}. 
Furthermore, some methods have been developed to treat the matter of binaries 
and internal gravity of them. 
For instance, in the post-Minkowskian approximation \cite{DD81a,DD81b}, 
the self gravity causes divergent behavior at the position of a point source, 
so that regularization schemes have been used \cite{BDDIM,DD81a}. 
Other method assumed that the neutron star is modeled by 
the spherical distribution of the perfect fluid \cite{GK,Kopejkin}.  
In these derivations
where the standard post-Newtonian approximation is used,
however, the weak field is assumed everywhere or everywhere except at the singular point. 
As a result, strictly speaking, the obtained equation of motion 
is applicable only to a object with weak internal gravity or 
to a  singular source whose relation to the real system is not 
clear. 
It is then necessary to extrapolate the equivalence principle 
to the extended objects with strong self gravity, in order to 
apply the post-Newtonian equation of motion derived from the standard
approach to neutron star binaries. 
Since such an extrapolation is not trivial at all, it must be shown 
within the framework of general relativity if it works. 

The equation of motion for bodies with strong self gravity 
has been studied by various authors; D'Eath developed an approach 
based on the asymptotic matching \cite{DEath}. 
However, this approach is based on a perturbation of black holes. 
Hence, there remain several problems to be solved in this scheme; 
how can we take into account the difference between neutron stars 
and black holes? How can we go to nonlinear perturbations? 
(Recall that the nonlinear perturbation is required for equal-mass
binaries even at the 1 PN order.)
Thorne and Hartle's derivation is also using the linear order 
perturbation \cite{TH}. 
However, there seem a great of difficulties in pushing their method 
forward to obtain a higher order equation of motion. 

In the present paper, we take other approach, the so-called point 
particle limit in order to treat the strong self gravity where 
the size of the component star is reduced to a point 
keeping the strength of the internal gravity\cite{Futamase85,Futamase87}. 
The method has been used  to derive only the equation of motion at the
Newtonian order and the quadrupole radiation reaction at 2.5PN
order for compact binaries.  
Here and henceforth, by the post-Newtonian approximation, we mean 
only the smallness of the ratio of the orbital velocity to 
the light velocity and the weakness of the gravitational field 
outside the compact objects, but not the weakness of the gravitational 
field inside these. 
As discussed above, we need to know higher  post-Newtonian 
orders up to, say, 4th post-Newtonian order to have an accurate 
prediction of the emitted waveform \cite{TN94}.
This is what we are aiming at. Namely, we wish to derive 
the equation of motion applicable to systems with strong internal
gravity such as {\it compact binaries} up to a sufficiently 
higher post-Newtonian orders using the point particle limit. 
As the first step, here we derive the first and 
(partly) second post-Newtonian equation of motion 
for compact binaries taking fully strong internal gravity into account. 
For the sake of the completeness, we shall show some details of the approach
which has been proposed by one of the present author\cite{Futamase87}. 
Our main contributions in this paper are to derive
the first post-Newtonian equation of motion, the lowest spin-orbit
coupling force, the lowest quadrupole-orbit coupling force
for relativistic compact binaries and a general momentum velocity
relation which was not discussed in \cite{Futamase87}.

This paper is organized as follows. 
Section II presents a general framework for the derivation; 
the equation of motion expressed with surface integrals and 
the point particle limit. 
In section III, we review the Newtonian-like equation of motion 
for compact binaries to show our basic calculation.
We derive the 1PN equation of motion in section IV, 
namely with some remarks on the definition of the mass and momentum in
this paper 
and take account of effects of the quadrupole moment and the spin of 
compact objects in section V. 
Section VI is devoted to the conclusion.
In appendix A we will give some comments on the derivation of the
components of the metric.
Some formulae are given in the appendix B.

Greek indices take from 0 to 3, and Latin indices from 1 to 3.

\section{General Formalism} 
\subsection{Newtonian scalings and point particle limit}

It is expected that the orbital motion of neutron star binaries
in the beginning of the inspiralling phase are 
governed mainly by  Newtonian gravity so that it is natural to 
formulate our approach based on the post-Newtonian approximation. 
For our purpose it turns out convenient to introduce a Newtonian 
dynamical time as \cite{FS83}
\begin{equation}
\tau=\epsilon t . 
\end{equation}
Then, the coordinate velocity is 
\begin{equation}
\frac{dx^i}{dt} = \epsilon \frac{dx^i}{d\tau} , 
\end{equation}
so that the velocity scales as $\epsilon$. Note that events with
different $\epsilon$ but at the same Newtonian dynamical time $\tau$
are at (roughly) the same phase in the orbit. This motivates us to 
define the post-Newtonian sequence by letting $\epsilon$ go to zero 
with $\tau$ staying constant. We call this limit, 
$\epsilon \to 0$, the Newtonian limit. 
In the binary whose mass and separation are denoted by $M$ and $L$ 
respectively, the balance of Newtonian forces is written as 
\begin{equation}
|\frac{dx^i}{dt}|^2 \sim M/L . 
\end{equation}
Hence, we find that $M$ scales as $\epsilon^2$ when we fix 
the separation $L$. 

In order to treat strong internal gravity we introduce the point
particle limit in the post-Newtonian sequence. 
Namely, denoting a size of a neutron star by $R$, we expect 
$M/R$ is of order of unity. 
To be so, $R$ must scale as $\epsilon^2$, which means that  
the density is of order $\epsilon^{-4}$. 
Hence, we define the body zones $B_A$ and the body zone coordinates as $
B_A = \{x^k | |x^k - z_A^k(\tau)| < \epsilon R_A\}$
for some $R_A$ and
\begin{equation} 
\alpha^{\underline{k}}_A = \epsilon^{-2}(x^k-z_A^k(\tau)) , 
\end{equation}
so that in the body zone coordinate, a size of a neutron star remains 
constant 
even as $\epsilon \to 0$. Here the underlined indices
refer to the body zone coordinate and $z^i_A(\tau)$, $A = 1,2$, are the
near zone coordinates of the representative points of two bodies.
There are three purposes to
define the body zone and the body zone coordinate. First, it is useful 
to define
physical quantities such as mass, spin with the  body zone coordinate, since
in this coordinate the body is nearly isolated. Second, as can be seen
later, we define a gravitational force on the object A
as a momentum flow per unit time
through the surface of the body zone of the object A. 
Thus we only need to know the metric at the boundary of the body zone
where the assumption of weak gravity is valid.
To deal with strong internal
gravity, this is a great advantage since we do not have to care about the
internal strong gravity. Third, the multipole moments of the body are
reduced to the higher order terms relative to monopole terms thanks to
the scaling of the body zone coordinate, as will be seen later. This is
a nice feature because in the inspiralling phase
it is a good approximation
that compact objects are
expressed as point particles with their multipoles as higher corrections.

In this paper  we adopt the initial value approach as in
\cite{Schutz80,Futamase87} and  we
choose two stationary uniformally rotating fluids with spatially compact support as
the initial data for the matter. Also, we assume that both objects
rotate slowly, that is, the velocities of the spinning motion of the bodies
scale as $\epsilon$.
It is easy to incorporate rapidly rotating stars into our formalism.
The result is that a coefficient $\epsilon^3$ appears instead of the
coefficient $\epsilon^4$ in the equation (\ref{SOcouplingforce}).
We note that, with this scaling, the spin-orbit
coupling
force appears at the 2PN order, which is different from the
usual result \cite{Damour,TH,Kidder}.
The initial data of the stress-energy tensor $T_A^{\mu\nu}$
of the body $A$ is
supposed to scale like 
\begin{eqnarray}
T_A^{\tau\tau}&=&\epsilon^2 T_A^{tt} \sim \epsilon^{-2} , \\
T_A^{\tau \underline{i}}&=&\epsilon^{-1} T_A^{ti} \sim \epsilon^{-4} , \\
T_A^{\underline{ij}}&=&\epsilon^{-4} T_A^{ij} \sim \epsilon^{-8} .  
\end{eqnarray}
Note that as a slowly rotating fluid, the object is supported by the
pressure. Also, note that the assumption that the densities of the
fluids scale as 
$\epsilon^{-4}$
are applied to the components in the near zone
coordinate,
not in the body zone coordinate.
The scalings of the stress energy
tensor of bodies on the initial hypersurface 
can be found in the Table \ref{TableOfScalings}.

 Before ending up this subsection, we
mention briefly 
the coordinate
transformation from the near zone coordinate to the Fermi normal
coordinate of the extended body 
which is needed to derive the spin-orbital coupling force in the
same form as the previous works  \cite{Damour,TH,Kidder}.

These transformation may be obtained order by order, and
at the lowest
order, we identify this transformation as the Galilei transformation,
so that we have
\begin{eqnarray}
T^{\tau\tau}_N &=& T^{\tau\tau}_A,
 \label{transformationTT} \\ 
T^{\tau i}_N &=& \epsilon^2 T^{\tau \underline{i}}_A +
 T^{\tau \tau}_A v_A^i, 
\label{transformationTI} \\
T^{ij}_N &=& T^{\tau\tau}_A v_A^i v_A^j +
 2 \epsilon^2 v_A^{(i} T_A^{\underline{j})\tau}
 + \epsilon^4 T_A^{\underline{ij}}.  
\label{transformationIJ}
\end{eqnarray}
where the subscript N 
means that these are components of the stress energy tensor
in the near zone coordinate. These equations are enough for
obtaining the 1PN equation of motion.

It is difficult, however, to obtain the transformation even at
the next order when we consider a body with strong internal gravity
(for the case of weak internal gravity, see e.g. \cite{BK,DSX} ).
Fortunately, it turns out in section V
that 
we do not
need to know an explicit form of the transformation
to derive the lowest spin-orbital
coupling force.

\subsection{A general form of the equation of motion}
As discussed above we need to know the metric at the boundary of the
body zone where it is assumed that the metric deviates
 slightly from the flat metric. 
Thus we define the small deviation from the flat metric $\eta^{\mu\nu}$ as 
\begin{equation} 
\bar h^{\mu\nu}=\eta^{\mu\nu}-\sqrt{-g}g^{\mu\nu} , 
\end{equation}
where we denote the determinant of the metric $g_{\mu\nu}$ by $g$.  

We choose the harmonic condition on the metric as 
\begin{equation}
\bar h^{\mu\nu}{}_{,\nu}=0 , 
\end{equation}
where the comma denotes the partial derivative. 
Then, the Einstein's equation is rewritten as 
\begin{equation}
\Box \bar h^{\mu\nu} = -16\pi \Lambda^{\mu\nu} , 
\end{equation}
where we defined 
\begin{eqnarray}
&&\Box = \eta^{\mu\nu}\pa_{\mu}\pa_{\nu} , \\
&&\Lambda^{\mu\nu} = \Theta^{\mu\nu}
+\chi^{\mu\nu\alpha\beta}{}_{,\alpha\beta} , \\
&&\Theta^{\mu\nu} = (-g) (T^{\mu\nu}+t^{\mu\nu}_{LL}) , \\
&&\chi^{\mu\nu\alpha\beta} = \frac{1}{16\pi} 
(\bar h^{\alpha\nu}\bar h^{\beta\mu}
-\bar h^{\alpha\beta}\bar h^{\mu\nu}) . 
\end{eqnarray}
Here, $T^{\mu\nu}$ and $t^{\mu\nu}_{LL}$ denote the stress-energy tensor 
and the Landau-Lifshitz pseudotensor \cite{LL}, respectively. 
The conservation law is expressed as 
\begin{equation}
\Lambda^{\mu\nu}{}_{,\nu}=0 . 
\label{conservation}
\end{equation}

The reduced Einstein's equation is solved formally as 
\begin{equation}
\bar h^{\mu\nu}=4 \int_{C(\tau, x^k; \epsilon)} d^3y 
\frac{\Lambda^{\mu\nu}_N(\tau-\epsilon|\vec x-\vec y|, y^k; \epsilon)} 
{|\vec x-\vec y|} ,  
\label{IntegratedREE}
\end{equation}
where $C(\tau, x^k; \epsilon)$ means the past light cone emanating 
{}from the event $(\tau, x^k)$. 
To clarify that $\Lambda^{\mu\nu}$ appearing
in the Eq. (\ref{IntegratedREE}) are components in the near zone coordinate
$(\tau,x^k)$, we added a subscript $N$ to $\Lambda^{\mu\nu}$.  
This equation must be solved iteratively to obtain an explicit form 
of the metric. We have ignored the homogeneous solution, since it is 
irrelevant to the dynamics of the binary at the order considered in
this paper\cite{Futamase83}.

Let us define the four momentum of the object A as 
the integral over the body zone $B_A$ like
\begin{equation}
P^{\mu}_A(\tau)= \epsilon^2 \int_{B_A} d^3\alpha_A
 \Lambda^{\tau \mu}_N(\tau,\alpha_A^{\underline{k}}; \epsilon) . 
\label{DefOfMom}
\end{equation}
Note that  
$\chi^{\tau\nu\alpha\beta}$ does not play any role up to 
the 2PN order. 
We can understand this from the fact that the lowest
order of the fields $\bar h^{\mu\nu}$ is $\epsilon^4$, which is seen in
the next section, and
the volume integrals
of $\chi^{\tau\nu\alpha\beta}\mbox{}_{,\alpha\beta}$ 
can be transformed into the surface integrals.

It should be noted that $P^{\mu}_A$ depends only on the time 
coordinate. We call the  $\mu=\tau$ and $i$ component of $P^{\mu}_A$
simply the energy and the momentum of the object $A$ respectively. 
Note that we have not specified $z^i_A(\tau)$ yet. 
We define the derivative which does not change the region $B_A$ as 
\begin{equation}
\frac{D}{D \tau}=\frac{\pa}{\pa \tau}+v_A^k \frac{\pa}{\pa x_k} ,   
\label{derivative}
\end{equation}
where we defined $v_A^k$ as 
\begin{equation}
v_A^k=\dot z^k_A(\tau), 
\end{equation}
and the dot denotes a temporal derivative. 
Note that the differential operator $D/D\tau$ differs essentially from
the usual Lagrange derivative $d/d\tau$. But since the body zones for each
body
remain unchanged 
(in the near zone
coordinate sense), these two operators are effectively the same
when they act on the integrals over the body zones.

Here, we define the near zone dipole moment of the object as 
\begin{equation}
D_A^i(\tau) = \epsilon^2 \int_{B_A} d^3\alpha_A \Lambda^{\tau\tau}_N
 (\tau,\alpha_A^{\underline{k}}; \epsilon)
 \alpha_A^{\underline{i}} . 
\end{equation}
To define the center of the mass $z^i_A(\tau)$ of the object A
it is required that $D^i_A$ takes a certain value. At the Newtonian
and post-Newtonian order, we will require that $D_A^i$ vanishes.
But to derive the spin-orbit coupling force in the usual form 
\cite{Damour,TH,Kidder}, care must
be taken when we define the center of the mass as can be seen later.
By taking the temporal derivative of $D^i_A$, we obtain 
\begin{equation}
P_A^i=P_A^{\tau} v_A^i+Q_A^i + \epsilon^2 \frac{d D_A^i}{d\tau}, 
\label{velocity-momentum}
\end{equation}
where we used the derivative defined by Eq. ($\ref{derivative}$).  
Here, we defined $Q_A^i$ as 
\begin{equation}
Q_A^i=\epsilon^{-4}
 \oint_{\pa B_A} dS_k \Lambda^{\tau k}_N
 \Bigl\{ x^i-z_A^i(\tau) \Bigr\} 
-\epsilon^{-4}
v_A^k \oint_{\pa B_A} dS_k
\Lambda^{\tau\tau}_N \Bigl\{ x^i-z_A^i(\tau) \Bigr\} , 
\label{Q}
\end{equation} 
where $\pa B_A$ denotes the surface of the sphere $B_A$. 
Equation ($\ref{velocity-momentum}$) is thought as the relation between 
the
{\it velocity}
and {\it momentum} if we make a certain choice   
of the near zone dipole moment $D_A^i$, that is, we define the center
of mass of the object A. 
It is worthwhile to mention that the velocity and the momentum are 
not proportional to each other, but $Q_A^i$ appears as its correction 
to be considered. 
Actually, $Q_A^i$ contributes to the equation of motion as shown in the 
section V.

Using the conservation law, Eq. ($\ref{conservation}$), 
we obtain  
\begin{equation}
\frac{d}{d\tau}P_A^{\mu} = -\epsilon^{-4}
 \oint_{\pa B_A} dS_k \Lambda^{k\mu}_N
+\epsilon^{-4}
v_A^k \oint_{\pa B_A} dS_k \Lambda^{\tau\mu}_N , 
\label{momentum}
\end{equation}
where we used Eq. ($\ref{derivative}$). 
This is an evolution equation for the four momentum $P_A^{\mu}(\tau)$.

In order to obtain the equation of motion, an evolution equation 
for the momentum must be changed into an evolution equation 
for the velocity. 
Inserting Eq. ($\ref{velocity-momentum}$) into the i-th component of 
Eq. ($\ref{momentum}$), 
we obtain the general form of the equation of motion as 
\begin{eqnarray}
P_A^{\tau}\frac{d}{d\tau}v_A^i =&&
 -\epsilon^{-4}
 \oint_{\pa B_A} dS_k \Lambda^{ki}_N
+ \epsilon^{-4}
v_A^k \oint_{\pa B_A} dS_k \Lambda^{\tau i}_N 
\nonumber\\
&&
 +\epsilon^{-4}
 v_A^i \Bigl( \oint_{\pa B_A} dS_k \Lambda^{k\tau}_N
-v_A^k \oint_{\pa B_A} dS_k \Lambda^{\tau\tau}_N \Bigr)
\nonumber\\
&&-\frac{d}{d\tau}Q_A^i  - \epsilon^2 \frac{d^2 D_A^i}{d \tau^2}.  
\label{generaleom}
\end{eqnarray} 
The above set of our basic equations 
(\ref{Q}) and 
($\ref{generaleom}$) consist only of the surface integrals when the
representative point of the body A, $z^i_A$ or equivalently, the value
of the near zone dipole moment $D_A^i$ is specified. 
Therefore, all we must do is to evaluate these surface integrals 
perturbatively. 
This is a great advantage of our formalism, since our evaluations 
can be done outside the strong self gravitational field.

\section{Newtonian equation of motion}

{}From the point of view of the point particle limit, 
the Newtonian order of the equation of motion for compact objects 
has been examined, where $Q_A^i$ was not considered
in
\cite{Futamase87}. 
However, since $Q_A^i$ is newly introduced in this paper,
we rederive the Newtonian equation of motion in this section.
We also intend the completeness and to show our basic calculation.

At the Newtonian order, we use equations (\ref{transformationTT}),
(\ref{transformationTI}) and (\ref{transformationIJ}).
The near zone dipole moment becomes
\begin{equation}
D_A^i(\tau) = \epsilon^2 \int_{B_A} d^3\alpha_A
 \Lambda^{\tau\tau}_A(\tau,\alpha_A^{\underline{k}}; \epsilon)
 \alpha_A^{\underline{i}} . 
\end{equation}
So, we define the center of mass of the object A as $z_A^i$, 
by requiring that the near zone dipole moment vanishes
\begin{equation}
D_A^i(\tau) = 0 . 
\end{equation}

Then metric up to $O(\epsilon^4)$ becomes
\cite{Futamase85}
\begin{eqnarray}
\bar h^{\tau\tau}&=&4\epsilon^4\sum_{A=1,2} \frac{m_A}{r_A}
+O(\epsilon^5) ,
\label{h4tt} \\
\bar h^{\tau i}&=&4\epsilon^4\sum_{A=1,2} \frac{J_A^i}{r_A}
+4\epsilon^4\sum_{A=1,2} \frac{m_A v_A^i}{r_A}+O(\epsilon^5) ,
\label{h4ti} \\
\bar h^{ij}&=&4\epsilon^2\sum_{A=1,2} \frac{Z_A^{ij}}{r_A}
+8\epsilon^4\sum_{A=1,2} \frac{v_A^{(i}J_A^{j)}}{r_A}
+4\epsilon^4\sum_{A=1,2} \frac{m_A v_A^i v_A^j}{r_A} \nonumber \\
\mbox{} &+&
 \epsilon^4 \sum_{A=1,2}\frac{m_A^2}{r_A^4}r_A^i r_A^j
-\epsilon^4 \frac{8 m_1 m_2}{r_{12}S}n_{12}^i n_{12}^j
\nonumber \\
\mbox{}&-&
8\epsilon^4 \left[\delta^{i}\mbox{}_{k}\delta^{j}\mbox{}_{l}
  - \frac{1}{2}\delta^{i j}\delta_{k l}\right]
\frac{m_1 m_2}{S^2}(\vec{n}_{12}-\vec{n}_{1})^{{\scriptscriptstyle (}k}
                   (\vec{n}_{12}+\vec{n}_{2})^{l{\scriptscriptstyle )}}
\nonumber \\
\mbox{}&+& O(\epsilon^5) ,
\end{eqnarray}
where $r_A^i = x^i - z_A^i$, $r_A = |\vec{r}_A|$,
$r_{12}^i = z_1^i - z^i_2$ and  $S = r_1 + r_2 + r_{12}$.
 We defined
\begin{eqnarray}
&&m_A=\lim_{\epsilon\to 0} \epsilon^2 \int_{B_A} d^3\alpha_A 
\Lambda^{\tau\tau}_A(\tau,\alpha_A^{\underline{k}}; \epsilon), \\
&&J_A^i=\lim_{\epsilon\to 0} \epsilon^4 \int_{B_A} d^3\alpha_A 
\Lambda^{\tau \underline{i}}_A(\tau,\alpha_A^{\underline{k}}; \epsilon), \\
&&Z_A^{ij}=\lim_{\epsilon\to 0} \epsilon^8 \int_{B_A} d^3\alpha_A 
\Lambda^{\underline{ij}}_A(\tau,\alpha_A^{\underline{k}}; \epsilon) . 
\end{eqnarray}
The $m_A$ is the ADM mass that the body $A$ would have
if the body $A$ were isolated.
By choosing the center of mass, $J_A^i$ vanishes. Furthermore, we assume 
that the fluids of the two bodies are (quasi-) stationary so that $Z_A^{ij}$
are of higher order, $\epsilon^6$
\cite{Futamase87}.
Note that at the Newtonian order, we have $P^{\tau}_A = m_A$ by virtue
of Eq. (\ref{transformationTT}). We also note that
the temporal derivative of the $P^{\tau}_A$
vanishes at the Newtonian order from Eqs.
(\ref{tLL4tt}) and (\ref{tLL4ti}). 
To obtain the Newtonian equation of motion, we need only the field
$\bar{h}^{\tau\tau}$ of order $\epsilon^4$.

Now, to evaluate the surface integrals we take 
\begin{eqnarray}
&&\vec r_1=\epsilon R_1 \vec n_1 , \\
&&\vec r_2=\vec r_{12}+\epsilon R_1 \vec n_1 , \\
&&\vec r_{12}=\vec z_1-\vec z_2 , 
\end{eqnarray}
where $\vec n_1$ denotes the spatial unit vector 
emanating from $\vec z_1$ and $\epsilon R_A$ is the radius of 
the sphere $B_A$. 
The pseudotensor at this order is 
\begin{equation}
{}_{(4)}[(-g)t^{ik}_{LL}]=\frac{1}{64\pi} 
\Bigl( {}_{(4)} \bar h^{\tau\tau,i} {}_{(4)} \bar h^{\tau\tau,k} 
-\frac12 \delta^{ik} {}_{(4)} \bar h^{\tau\tau,l} 
{}_{(4)} \bar h^{\tau\tau}{}_{,l} \Bigr) , 
\end{equation}
where $(n)$ means the order of $\epsilon^{n}$. 
We evaluate 
\begin{eqnarray}
\oint_{\pa B_1} dS_k {}_{(4)}[(-g)t^{ik}_{LL}]
&=&\frac{1}{64\pi} \Bigl( \delta^i_l \delta^k_m -\frac12 
\delta^{ik}\delta_{lm} \Bigr)
\sum_{A=1, 2}\sum_{B=1, 2} \oint_{\pa B_1} dS_k 
\frac{m_A m_B r_A^l r_B^m}{r_A^3 r_B^3} , \nonumber \\
\mbox{} &=&\frac{m_1 m_2}{r_{12}^2}n_{12}^i ,  
\end{eqnarray}
where we defined $n_{12}^i$ as 
\begin{equation}
\vec n_{12}=\frac{\vec r_{12}}{|\vec r_{12}|} . 
\end{equation}
Hence, we obtain 
\begin{equation}
\frac{dP_1^i}{d\tau}=-\frac{m_1 m_2}{r_{12}^2}n_{12}^i . 
\label{Ndotmomentum}
\end{equation}

At the Newtonian order, we have $Q_A^i = 0$ from Eqs.
(\ref{tLL4tt}) and (\ref{tLL4ti})
, so we have  the Newtonian
momentum velocity relation, $P_A^i = m_A v_A^i$.
Hence we obtain the Newtonian equation of motion for relativistic
compact binaries.

\section{The first post-Newtonian equation of motion}

In this section, we shall derive the 1PN equation of motion. At this order
$T_N^{\tau\tau}$ is transformed as
\begin{eqnarray} 
T_N^{\tau\tau} = (\Gamma_A)^2 T_A^{\tau\tau} +
2 \epsilon^4 v_A^i T_A^{\underline{i}\tau} + O(\epsilon^6),
\label{transform1PNtt}
\end{eqnarray}
where $\Gamma_A$ is a transformation coefficient whose explicit
expression  is irrelevant for the purpose in this paper. 
The second term of the above equation does not contribute to
the near zone dipole moment when we require that $J_A^i$ vanishes,
so
the condition $D_A^i = 0$ defines the same center of the mass as in the
Newtonian case.
However, we must take into
account this transformation when we consider the spin-orbit coupling
force as seen in the next section.

As for $Q_A^i$, this term  appears from the 0.5PN order and depends
on the size of the body zone.
However up to  the 0.5PN order, $Q_A^i$
does not affect the equation of motion. We will show this explicitly in the
following two subsections.

\subsection{The boundary dependent terms}

Before deriving the 1PN equation of motion,
we present some comments
on the mass and momentum defined in this paper, which are relevant to
the cancellation between terms depending
on the size of the body zone; our definition
depends on the size of the body zone  since 
the pseudotensor has non-compact support. 
The boundary of the body zone is given by hand so that 
the mass must depend on the size of the body zone. 
Thus, we may split the mass as 
\begin{equation}
m_A=\bar m_A+\tilde m_A, 
\end{equation}   
where $\bar m_A$ does not depend on the boundary size but 
$\tilde m_A$ does.  
In order to perform this splitting, we use the fact that 
the integral of $\Lambda^{\tau\tau}_N$ over the whole space $W$ does not 
depend on the size of the body zone, say $\epsilon R_1$. 
In fact, from the relevant expression of the pseudotensor, the
equations (\ref{tLL6tt})  in the appendix B,
we obtain at the lowest order  
\begin{eqnarray}
\epsilon^2 \int_{W/B_1} d^3\alpha
 \epsilon^6 \mbox{}_{(6)}\Lambda^{\tau\tau}_N
&=& \epsilon^{2}\int_{W/B_1} d^3y \mbox{}_{(6)}\Lambda^{\tau\tau}_N
\nonumber \\
&=& -\epsilon^2\frac78
 \int_{\epsilon R_1}^{\infty} y_1^2 d|y_1| \oint d\Omega_1 
\Bigl( \frac{m_1}{|y_1|}\Bigr)_{,k} 
\Bigl( \frac{m_1}{|y_1|}\Bigr)_{,k}
\nonumber \\
&+&  ({\rm terms\,independent\,of\,R_1})
\nonumber  \\
&=&-\epsilon^2 \frac72 \frac{m_1^2}{\epsilon R_1}
 + ({\rm terms\,independent\,of\,R_1}),  
\nonumber 
\end{eqnarray}
where $W/B_1$ means the spatial domain obtained by subtracting 
$B_1$ from $W$. 
Since the total energy must not depend on the size of the body zone,
we find 
\begin{equation}
\tilde m_A= \epsilon^2 \frac72 \frac{m_A^2}{\epsilon R_A}  
 + O(\epsilon^2).
\end{equation} 
In the similar manner, from Eq. (\ref{tLL6ti}) we split $P_A^i$ into  
$P_A^i=\bar P_A^i+\tilde P_A^i$  
and obtain 
\begin{equation}
\tilde P_A^i=\epsilon^2 \frac{11}{3} \frac{m_A^2 v_A^i}{\epsilon R_A}
 +O(\epsilon^2).
\label{TildeMom} 
\end{equation}
Furthermore, we evaluate each integral in Eq. ($\ref{Q}$) as
\begin{equation}
\epsilon^{-4}
 \oint_{\pa B_1} dS_k
 \mbox{}_{(6)}\Lambda^{\tau\tau}_N \Bigl\{ x^i-z_1^i(\tau) \Bigr\} 
=- \epsilon^2 \frac{7m_1^2}{6\epsilon R_1}
\label{FPNToHPN}, 
\end{equation}
\begin{equation}
\epsilon^{-4}
 \oint_{\pa B_1} dS_k
 \mbox{}_{(6)}\Lambda^{\tau k}_N \Bigl\{ x^i-z_1^i(\tau) \Bigr\} 
=-\epsilon^2\frac{m_1^2 v_1^i}{\epsilon R_1}. 
\end{equation}
Hence, we obtain
$\bar{Q}_A^i = O(\epsilon^4)$ and 
\begin{equation}
\tilde{Q}_A^i=\epsilon^2 \frac{m_A^2 v^i_A}{6 \epsilon R_A}+O(\epsilon^2),  
\end{equation}
where we split $Q_A^i$ like $m_A$.
Thus we have
\begin{equation}
 \bar P_A^i = \bar m_A v^i_A + O(\epsilon^2).
\label{renormalizedMVR}
\end{equation}
This is the momentum velocity relation independent of the size of the
boundary. Note that this relation is valid up to the 0.5PN order, since
$Q_A^i$ and all the terms which depend on the size of the body zone
are evaluated with the 1PN pseudotensor. There are other 
terms which depend on the size of the body zone. These  
come from the integrals which
are formally of the 2PN and 2.5PN order
(e.g. $\mbox{}_{(8)}\Lambda^{\tau\tau}_N$
in Eq. (\ref{FPNToHPN}) )
and, at first sight, contribute to the 1PN
equation of motion. We will discuss this issue in the forthcoming
paper where we will derive the 2.5PN equation of motion.

Since the boundary-dependent term of the mass is of order $\epsilon$,
when we pay attention only to the Newtonian order, we
can rewrite the Newtonian equation of motion as
\begin{equation} 
\bar{m}_1 \frac{d v_1^i}{d\tau}
 = 
 - \frac{\bar{m}_1\bar{m}_2}{r_{12}^2}n_{12}^i.
\label{renormalizedNEOM}
\end{equation}

\subsection{The first post-Newtonian equation of motion}

In this subsection we derive the 1PN  equation of motion
paying attention to  the cancellation between terms dependent on the size of
the body zone.

First we consider the temporal component of the four momentum 
at the 1PN order. 
{}From Eqs. (\ref{tLL6tt}), (\ref{tLL6ti}), (\ref{h4tt}) and
(\ref{h4ti}), we obtain 
\begin{eqnarray}
\frac{dP_1^{\tau}}{d\tau}&=&-\epsilon^{-4}
 \oint_{\pa B_1} dS_k \epsilon^6 {}_{(6)}\Lambda^{k\tau}_N
+\epsilon^{-4}
v_1^k \oint_{\pa B_1} dS_k \epsilon^6 {}_{(6)} \Lambda^{\tau\tau}_N
\nonumber \\
&=&-\epsilon^2\frac{\bar{m}_1\bar{m}_2}{r_{12}^3}
 \Bigl( 4(\vec r_{12}\cdot\vec v_1)
-3(\vec r_{12}\cdot\vec v_2) \Bigr) , 
\end{eqnarray}
where we used the following integral formulae 
\begin{eqnarray}
\frac{1}{16\pi} \oint_{\pa B_1} dS_m {}_{(4)}\bar h^{\tau\tau,i} 
{}_{(4)}\bar h^{\tau\tau,j}
=\frac{4\bar{m}_1\bar{m}_2\delta^i_m r_{12}^j}{3 r_{12}^3}
+\frac{4\bar{m}_1\bar{m}_2\delta^j_m r_{12}^i}{3 r_{12}^3} , 
\end{eqnarray}
\begin{eqnarray}
\frac{1}{16\pi} \oint_{\pa B_1} dS_m {}_{(4)}\bar h^{\tau\tau,i} 
{}_{(4)}\bar h^{\tau j,k}
=\frac{4\bar{m}_1\bar{m}_2 v_1^j \delta^k_m r_{12}^i}{3 r_{12}^3}
+\frac{4\bar{m}_1\bar{m}_2 v_2^j \delta^i_m r_{12}^k}{3 r_{12}^3} . 
\end{eqnarray}
Using the Newtonian equation of motion, this is rewritten as 
\begin{equation}
\frac{dP_1^{\tau}}{d\tau}=\epsilon^2 \bar{m}_1 \frac{d}{d\tau} 
\Bigl( \frac12 v_1^2+\frac{3\bar{m}_2}{r_{12}} \Bigr) . 
\end{equation}
Hence, we obtain 
\begin{equation}
P_1^{\tau}= m_1 \Bigl[ 1+\epsilon^2 \Bigl( \frac12 v_1^2 
+\frac{3\bar{m}_2}{r_{12}} \Bigr) \Bigr] +O(\epsilon^3) , 
\label{1PNmass}
\end{equation}
where we renormalized the integration constant into $m_1$ and 
replaced $\bar{m}_1$ by $m_1$ at the $\epsilon^2$ order 
because the difference is of order $\epsilon$ higher.
We split $P_A^{\tau}$ into
the boundary independent part
$\bar{P}_A^{\tau}$
and the boundary dependent part
$\tilde{P}_A^{\tau}$.
Up to the 0.5PN order $\tilde{P}_A^{\tau} = \tilde{m}_A$.

Now we turn to the equation of motion. The relevant expressions for
the pseudotensor and the metric components
are given
in the appendix B.
The surface integrals  appearing
here
are as follows
\cite{InvalidExpression};
\begin{eqnarray}
\oint_{\pa B_1} dS_k {}_{(4)}\Bigl( -g t_{LL}^{ik}\Bigr)
=&&\frac{\bar{P}_1^{\tau}\bar{P}_2^{\tau}}{r_{12}^2}n_{12}^i , 
\label{InvalidExpressionA} \\
\oint_{\pa B_1} dS_k {}_{(6)}\Bigl( -g t_{LL}^{ik}\Bigr)
=&&
\frac{\bar{m}_1\bar{m}_2}{r_{12}^2}n_{12}^i
\Bigl( -\frac32(\vec v_2\cdot\vec n_{12})^2 
+\frac32 v_2^2-\frac83 (\vec v_1\cdot\vec v_2) -\frac{8\bar{m}_1}{r_{12}} 
-4\frac{\bar{m}_2}{r_{12}} \Bigr) \nonumber \\  
&&
-\frac{\bar{m}_1\bar{m}_2}{r_{12}^2} 
(\vec{v}_2\cdot\vec{n}_{12})v_1^i
+\frac{\bar{m}_1\bar{m}_2}{r_{12}^2}
\Bigl(\frac83(\vec{v}_1\cdot\vec{n}_{12})
- 3(\vec{v}_2 \cdot \vec{n}_{12}) 
\Bigr)v_2^i   \nonumber \\
&&- \epsilon^2
 \frac{11 \bar{m}_1^2 \bar{m}_2}{3 \epsilon R_1 r_{12}^2} n_{12}^i,
\label{IntegratedtLL6ik} \\
v_1^k \oint_{\pa B_1} dS_k {}_{(6)}\Bigl( -g t_{LL}^{i\tau} \Bigr) 
=&&\frac{\bar{m}_1\bar{m}_2}{r_{12}^2} n_{12}^i 
\Bigl( \frac43 (\vec v_1\cdot\vec v_2)-v_1^2 \Bigr) 
-\frac{\bar{m}_1\bar{m}_2}{r_{12}^2}(\vec v_2\cdot\vec n_{12}) v_1^i 
\nonumber\\
&&-\frac{4 \bar{m}_1\bar{m}_2}{3 r_{12}^2}
(\vec n_{12}\cdot\vec v_1) v_2^i . 
\end{eqnarray}
We note that the last term in Eq. (\ref{IntegratedtLL6ik}) precisely 
cancels out the time derivative of  $\tilde{P}_A^i$, Eq. (\ref{TildeMom})
(or equivalently, $\tilde{m}_A v_A^i + Q_A^i$) when we use the
Newtonian equation of motion (\ref{renormalizedNEOM}).

Inserting these results and Eq. ($\ref{1PNmass}$) into 
Eq. ($\ref{generaleom}$) and using Eq. (\ref{renormalizedMVR}), we obtain 
\begin{eqnarray}
\bar{m}_1 \frac{dv_1^i}{d\tau}=&&
 -\frac{\bar{m}_1\bar{m}_2}{r_{12}^2}n_{12}^i 
\nonumber\\
&&+\epsilon^2 \frac{\bar{m}_1\bar{m}_2}{r_{12}^2} \Bigl[ n_{12}^i 
\Bigl( -v_1^2-2v_2^2+\frac32 (\vec v_2\cdot\vec n_{12})^2 
+4(\vec v_1\cdot\vec v_2)+\frac{5\bar{m}_1}{r_{12}}
+\frac{4\bar{m}_2}{r_{12}} \Bigr)
\nonumber\\
&&~~~~~~~~+V^i \Bigl( 4(\vec v_1\cdot\vec n_{12}) 
-3(\vec v_2\cdot\vec n_{12}) \Bigr) \Bigr] , 
\label{1PNEOMFinal}
\end{eqnarray}
where we defined the relative velocity as 
\begin{equation}
\vec V=\vec v_1-\vec v_2 . 
\end{equation}

Thus, it has been shown that the terms dependent on the
size of the body zone do not affect the 1 PN equation of  
motion, since parts proportional to $R_A^{-1}$ cancel out totally.

The equation (\ref{1PNEOMFinal}) 
takes the same form as those in previous works 
\cite{BDDIM,DD81a}. 
Hence, the applicability of the 1PN equation of motion 
has been extended to compact objects.

\section{The equation of motion with effects of the quadrupole moment 
and the spin}

Now we consider the forces associated with the spins and quadrupole
moments of the component stars in the equation of motion which 
appear at the 2PN order in our ordering. 

First, we must derive the metric components which depend on the spin and the quadrupole
moment.  
The calculations are straightforward except for the spin part of the
field ${}_{(8)}\bar h^{\tau\tau}$. 
{}From the equation (\ref{transform1PNtt}) and
the requirement that $J_A^i$ vanishes, we can calculate this term as 
\begin{eqnarray}
\bar h^{\tau\tau} &=&
 4 \epsilon^6\sum_{A=1,2}
 \int_{B_A} d^3\alpha_A
 \frac{\Lambda^{\tau\tau}_N
 (\tau-\epsilon |\vec{r}_A - \epsilon^2 \vec{\alpha}_A|,
 \vec{\alpha}_A;\epsilon )}
 {|\vec{r}_A - \epsilon^2 \vec{\alpha}_A|}
 + 4\int_{C/B} d^3y \frac{\Lambda^{\tau\tau}_N
 (\tau-\epsilon|\vec{x} - \vec{y}|,
 \vec{y};\epsilon )}
 {|\vec{x} - \vec{y}|}  \nonumber \\
\mbox{} &=& 
 4 \epsilon^4\sum_{A=1,2}\left[\frac{1}{r_A}
 \int_{B_A}d^3\alpha_A
 \epsilon^2 \Lambda^{\tau\tau}_N(\tau, \vec{\alpha}_A;\epsilon )
+ \frac{\epsilon^2 r_A^i}{r_A^3}
 \int_{B_A} d^3\alpha_A
 \epsilon^2 \alpha_A^{\underline{i}}
 \Lambda^{\tau\tau}_N(\tau, \vec{\alpha}_A;\epsilon )
						 \right] \nonumber \\
\mbox{} &+& ({\rm terms\, irrelevant\, to\, the\, spin\, and\,
higher\, order\, terms\, than\, 2PN})  \nonumber \\
\mbox{} &=& 
 4 \epsilon^4\sum_{A=1,2}\left[\frac{\bar{P}^{\tau}_A}{r_A}
+ \epsilon^2 \frac{r_A^i}{r_A^3}(d_A^i + \epsilon^2 M_A^{ik} v_A^k)
						 \right] \nonumber \\
\mbox{} &+& ({\rm terms\, irrelevant\, to\, the\, spin\, and\,
higher\, order\, terms\, than\, 2PN}),
\end{eqnarray}
where in the last equality, we used the equation $(2.19)$
in \cite{Futamase87}.
We defined $d_A^i$ and the spin tensor $M_A^{ij}$ of the body A as 
\begin{eqnarray}
d_A^i &=& \lim_{\epsilon\to 0} \epsilon^2 \int_{B_A} d^3\alpha_A 
\Gamma_A^2 \alpha^{\underline{i}}_A\Lambda^{\tau\tau}_A, \\
M_A^{ij} &=& \lim_{\epsilon\to 0} 2 \epsilon^4 \int_{B_A} d^3\alpha_A 
\alpha^{\underline{[i}}_A\Lambda^{\underline{j}]\tau}_A.
\end{eqnarray}
Recalling that $\Lambda^{\tau\tau}_A$ is a component of a tensor in
the Fermi normal
coordinate, we define the center of mass by requiring that $d_A^i$,
instead of $D_A^i$, vanishes. 

It is important to note that the above choice of the center of the mass
leads us to the following velocity-momentum relation;
\begin{eqnarray}
P_1^i &=& P_1^{\tau} v_1^i + Q_1^i
 - \epsilon^4 \frac{m_2 M_1^{ij} n_{12}^j}{r_{12}^2}.
\label{velocity-momentumWithSpin}
\end{eqnarray}
Here we used the fact that the temporal derivative of the spin $M_A^{ij}$
is of order $\epsilon^2$
\cite{Futamase87}.

The required metric components are expressed as 
\begin{eqnarray}
{}_{(8)}\bar h^{\tau\tau}&=&6\sum_{A=1,2} \frac{r_A^kr_A^l }{r_A^5} 
\hat I_A^{kl} +4\sum_{A=1,2} \frac{r_A^k}{r_A^3} M_A^{ki} v_A^{i} , 
\nonumber\\
{}_{(6)}\bar h^{\tau i}&=&2\sum_{A=1,2} \frac{r_A^k}{r_A^3} M_A^{ki} , 
\nonumber\\
{}_{(6)}\bar h^{ij}&=&4\sum_{A=1,2} \frac{r_A^k}{r_A^3} M_A^{k(i} v_A^{j)} , 
\nonumber
\end{eqnarray}
where $\hat I_A^{ij}$ denotes the tracefree tensor
for the reduced quadrupole moment 
of the body $A$ defined as
\begin{eqnarray}
\hat I_A^{ij} &=& \lim_{\epsilon\to 0} \epsilon^2 \int_{B_A} d^3\alpha_A 
\Lambda^{\tau\tau}_A
(\alpha^{\underline{i}}_A\alpha^{\underline{j}}_A -
\frac{\delta^{ij}}{3}|\vec{\alpha}_A|^2). 
\end{eqnarray}

The spin vector $S_A^i$ is related with $M_A^{ij}$ as 
\begin{equation}
S_A^i=\frac{1}{2}\epsilon_{ijk} M_A^{jk} , 
\end{equation}
where $\epsilon_{ijk}$ means the spatially unit antisymmetric 
symbol.

The pseudotensor needed in this section is expressed as the equations
(\ref{tLL8tt}), (\ref{tLL8ti}) and  (\ref{tLL8ij}) in the appendix B.

\subsection{Quadrupole-orbital couplings}
First we calculate the contribution from quadrupole moments. 
The surface integrals that we need to evaluate are only 
\begin{equation}
\oint_{\pa B_1} dS_k {}_{(8)}\Bigl( -g t_{LL}^{ik} \Bigr) 
=- \frac{3}{2r_{12}^4} 
\Bigl( \bar{m}_1\hat I_2^{kl}+\bar{m}_2\hat I_1^{kl} \Bigr) 
\bigl( 2\delta^{il}n_{12}^k - 5n_{12}^i n_{12}^kn_{12}^l \Bigr) .  
\end{equation}
{}From Eq. ($\ref{generaleom}$), we obtain immediately the result:
\begin{equation}
\bar{m}_1 \frac{dv_1^i}{d\tau}=\epsilon^4 \frac{3}{2r_{12}^4} 
\Bigl(\bar{m}_1\hat I_2^{kl}+\bar{m}_2\hat I_1^{kl}\Bigr) 
\Bigl(2\delta^{il}n_{12}^k - 5n_{12}^i n_{12}^kn_{12}^l\Bigr) . 
\label{tidaleffect}
\end{equation}
Here we omitted the Newtonian and
the 1PN force and the spin-orbit coupling force.
Note 
that we use $\bar{m}_A$ instead of $P_A^{\tau}$ in the left hand side
of the equation (\ref{generaleom}) here since
the difference between $P_A^{\tau}$ and $\bar{m}_A$
does not affect the quadrupole-orbital coupling.
This is because
{}from the equations (\ref{tLL8tt}) and (\ref{tLL8ti}) the equation
of evolution for time component $P_A^{\tau}$, Eq. (\ref{momentum}),
does not contain the quadrupole moment at the 2PN order,
hence $P_A^{\tau}$ does not contain the quadrupole moment.

This equation, (\ref{tidaleffect}),
has the same form as the Newtonian equation of motion 
for the so-called tidal force. 
Therefore, it has been shown that the above equation is valid even 
for the compact objects. 

\subsection{spin-orbital couplings} 
Next we turn our attention to the spin-orbital coupling in the equation
of motion. The relevant surface integrals are as follows:  
\begin{eqnarray}
\oint_{\pa B_1} dS_k {}_{(8)}\Bigl( -g t_{LL}^{\tau i} \Bigr) 
=&&\epsilon^{-3}\frac{2\bar{m}_1}{3R_1^3}M_1^{ik} 
- \frac{2\bar{m}_1}{3 r_{12}^3} 
\Bigl( M_2^{il}\Delta^{lk}+M_2^{lk}\Delta^{li} \Bigr) 
\nonumber\\
&&- \frac{2\bar{m}_2}{15r_{12}^3} 
\Bigl( M_1^{il}\Delta^{lk}+3 M_1^{lk}\Delta^{li} \Bigr) , 
\\
\oint_{\pa B_1} dS_k {}_{(8)}\Bigl( -g t_{LL}^{ki} \Bigr) 
=&&\epsilon^{-3}\frac{2\bar{m}_1}{3R_1^3}M_1^{ik}v_1^k  
+\frac{\bar{m}_1}{r_{12}^3} \Bigl( \frac43 v_1^k-2v_2^k \Bigr) 
(M_2^{il}\Delta^{lk}+M_2^{lk}\Delta^{il}) 
\nonumber\\
&&+\frac{\bar{m}_2}{r_{12}^3} \Bigl[ v_1^k \Bigl( 
\frac{6}{5}M_1^{il}\Delta^{lk}+\frac{8}{5}M_1^{lk}\Delta^{li} \Bigr) 
\nonumber\\
&&~~~~~~~~~~~+v_2^k \Bigl( -\frac43 M_1^{il}\Delta^{lk} 
-2M_1^{lk}\Delta^{li} \Bigr) \Bigr] , 
\end{eqnarray}
where we defined 
\begin{equation}
\Delta^{ij}=\delta^{ij}-3n_{12}^i n_{12}^j . 
\end{equation}
We also need to evaluate the surface integrals in $Q_1^i$ which are 
\begin{eqnarray}
\oint_{\pa B_1} dS_k {}_{(8)}\Bigl( -g t_{LL}^{\tau k} \Bigr) 
\Bigl\{ x^i-z_1^i(\tau) \Bigr\} =
&&\frac{2\bar{m}_2}{3r_{12}^2} M_1^{ik} n_{12}^k. 
\end{eqnarray}
Then  we obtain 
\begin{equation}
Q_1^i=\epsilon^4\frac{2\bar{m}_2}{3r_{12}^2} M_1^{ik}n_{12}^k . 
\end{equation}
Note that the surface integral of
${}_{(8)}\Bigl( -g t_{LL}^{\tau\tau} \Bigr)$
does not depend on the spins of the bodies.

Inserting the above results into Eq. ($\ref{generaleom}$) and recalling
the velocity-momentum relation (\ref{velocity-momentumWithSpin}),
we obtain 
\begin{eqnarray}
\bar{m}_1 \frac{dv_1^i}{d\tau} =
-\epsilon^4 \frac{V^k}{r_{12}^3} 
\Bigl[ \Bigl( 2 \bar{m}_1 M_2^{il}+ \bar{m}_2 M_1^{il} \Bigr) \Delta^{lk} 
+2 \Bigl( \bar{m}_1 M_2^{lk}+\bar{m}_2 M_1^{lk} \Bigr) \Delta^{li} \Bigr], 
\label{SOcouplingforce}
\end{eqnarray}
where we omitted irrelevant terms to the spin-orbit coupling force.
Note that the difference between $P_A^{\tau}$ and $\bar{m}_A$ has no influence
on the spin-orbital coupling because of the same reason as in
the quadrupole-orbital coupling.

Here, it is worthwhile to mention that the terms  
proportional to $R_1^{-3}$ cancel out 
totally in the equation of motion as we expect. This equation is
exactly  the same as the equation in \cite{Damour,TH}.

\section{Conclusion}
Based on the point particle limit, we derived the equation of motion 
not only at the first post-Newtonian (1PN) order but also with effects of 
the quadrupole moment and the spin of compact objects at the 2PN order. 
It is worthwhile to mention why these effects are of the 2PN order,
though they are usually of the Newtonian and 
the 1PN order. 
In the point particle limit, the ratio of the size of compact objects 
to the separation of the binary is taken as an expansion parameter, 
which is equal to $\epsilon^2$. 
Hence, the quadrupole moment and the spin become of higher orders 
in the ratio of the radius to the separation. Also, we assume
that the velocity of the spinning motion of the body scales as $\epsilon$. 
Therefore, their effects appear at the 2PN order. 

As for the spin-orbit coupling, we must carefully define
the center of the mass of the object to have the same
expression as the previous works 
\cite{Damour,TH,Kidder}. It is well-known that the form of the
spin-orbit coupling force depends on the definition of the center of
the mass of the object \cite{Kidder}.

We emphasize that the definitions of the mass, spin, quadrupole moment
of objects include the effect of the {\it strong} self gravity
explicitly (for the case of the weak internal gravity, e.g. 
\cite{BK,DSX}). 
Also we
note that
we did not need any regularization schemes.  
As a result, we have shown that the post-Newtonian equation of motion 
can be applied to compact binaries with strong internal gravity when we use the
suitably defined mass, spin and quadrupole moment up to 
the order studied  
in this paper. 
We are pushing the present scheme ahead to complete the 2nd and 2.5 
post-Newtonian equation of motion for the strong self-gravitating
objects which will be investigated in the near future. 
It is also very interesting to see if the present scheme works for the 3rd 
post-Newtonian order. We hope to tackle this problem also in the near future.

\section*{Acknowledgments}

H.A. would like to thank Bernard F. Schutz for his hospitality 
at the Albert-Einstein-Institute, where a part of this work was done.
The authors are grateful to Masaru Shibata for his careful reading
of our manuscript and for his suggestion.
This work was supported in part by the Japanese Grant-in-Aid 
for Scientific Research from the Ministry of Education, Science 
and Culture, No.11740130 (H.A.).
Y.I. would like to acknowledge the support from
JSPS Research Fellowships for Young Scientists.

\appendix

\section{Comments on the derivation of the metric}

The mass and four momentum are defined as the volume integral of
$\Lambda^{\mu\nu}$ in the body zone so that these depend on the
size of the body zone. On the other hand the integrated form of the
reduced Einstein equation (\ref{IntegratedREE})
has no information of the size of the body zone,
 so we expect that the metric components do not depend on it.
We here show that the field
$\bar{h}^{\tau\tau}$ is really  boundary-independent
up to the 0.5PN order.

{}From Eq. (\ref{IntegratedREE}), we have up to the 1PN order
\begin{eqnarray}
\bar h^{\tau\tau} &=&
 4 \epsilon^4 \sum_{A = 1,2} \frac{P^{\tau}_A}{r_A} +
4\epsilon^6 \int_{C/B} d^3y 
\frac{\mbox{}_{(6)}\Lambda^{\tau\tau}_N(\tau, y^k)} 
{|\vec x-\vec y|}
+ O(\epsilon^7). \label{BoundaryDependent}   
\end{eqnarray}

We can evaluate explicitly
the boundary-dependent term arising from the non-compact
 source, i.e. the last term in the above equation.

{}From the Eqs. (\ref{tLL6tt}) and (\ref{h4tt}),
\begin{eqnarray} 
\int_{C/B} d^3y 
\frac{\mbox{}_{(6)}\Lambda^{\tau\tau}_N(\tau, y^k)} 
{|\vec x-\vec y|} &=&
- \frac{7}{8 \pi} \sum_{A = 1,2}
\int_{C/B} d^3y \frac{m_A^2}{|\vec{y}_A|^4 |\vec{x}-\vec{y}|} +
({\rm irrelevant \, terms}).
\end{eqnarray}

Then it is easily verified that
\begin{eqnarray} 
\int_{C/B} d^3y \frac{1}{|\vec{y}_1|^4 |\vec{x}-\vec{y}|}
&=& \int^{r_1}_{\epsilon R_1}
d^3y_1\frac{1}{|\vec{y}_1|^4 |\vec{r}_1-\vec{y}_1|}
+ \int_{r_1}^{\infty}
d^3y_1\frac{1}{|\vec{y}_1|^4 |\vec{r}_1-\vec{y}_1|}
+ ({\rm higer \, order \, terms})\nonumber \\
&=&
\frac{4 \pi}{r_1}\left(\frac{1}{\epsilon R_1}- \frac{1}{2 r_1}\right) 
+ ({\rm higer \, order \, terms}).\
\end{eqnarray}

Inserting the above result into Eq. (\ref{BoundaryDependent}) we obtain
\begin{eqnarray}
\bar{h}^{\tau\tau} &=&
 4\epsilon^4 \sum_{A} \frac{1}{r_A}\left(P_A^{\tau}
- \epsilon^2 \frac{7 m_A^2}{2\epsilon R_A}
						 \right)
+ O(\epsilon^6).
\end{eqnarray} 

Let us  split $P_A^{\tau}$ into $\bar{P}_A^{\tau}$
and $\tilde{P}_A^{\tau}$ as
$P^{\tau}_A = \bar{P}^{\tau}_A + \tilde{P}_A^{\tau}$
where $\tilde{P}_A^{\tau}$ depends on the size of
the boundary while the $\bar{P}_A^{\tau}$ does not. Then  
{}from the equation (\ref{1PNmass})
we have relations like
$P_A^{\tau} = m_A$, $\bar{P}_A^{\tau} = \bar{m}_A$ and
$\tilde{P}_A^{\tau} = \tilde{m}_A$
up to the 0.5PN order. 
In section IV we obtained the following relation  
$$\tilde{m}_A = \epsilon^2 \frac{7 m_A^2}{2\epsilon R_A}.$$ 
Thus  it is shown  that up to the 0.5PN order
the boundary-dependent term in $P_A^{\tau}$, i.e. $\tilde{P}_A^{\tau}$
cancels out the term arising from the non-compact source.
We can write the metric up to the 0.5PN order as 
\begin{eqnarray}
\bar{h}^{\tau\tau} &=&
 4\epsilon^4 \sum_{A} \frac{\bar{P}_A^{\tau}}{r_A}
+ O(\epsilon^6).
\end{eqnarray}

\section{The metric and  the pseudotensor expanded in  $\epsilon$}

For convenience, we collect some formulae here.
First, we show the components of the
Landau-Lifshitz pseudotensor expanded in
$\epsilon$ up to an order relevant in this paper. 
The lowest order in the pseudotensor is $\epsilon^4$
since the lowest order of the near zone field $\bar h^{\mu\nu}$
is $\epsilon^4$ as shown later.
The components at the fifth order are identically zero, while
the components at the seventh order vanish because
${}_{(5)}\bar h^{\mu\nu}$ and ${}_{(7)}\bar h^{\tau\tau}$ depend only
on time $\tau$ \cite{Futamase87}. 
The required orders of the gravitational field and the Landau-Lifshitz
pseudotensor to obtain a certain order equation of motion 
are summarized in the Tables  
\ref{TableOfGF} and \ref{TableOftLL}.

 $(\quad)$ and $[\quad]$ mean the symmetrization and the
antisymmetrization respectively. 

$O(\epsilon^4)$ 
	\begin{eqnarray}
       \mbox{}_{(4)}[-16 \pi g t_{LL}^{\tau \tau}] &=& 0,
\label{tLL4tt} \\
 	    \mbox{}_{(4)}[-16 \pi g t_{LL}^{\tau i}] &=& 0,
\label{tLL4ti} \\	
	  \mbox{}_{(4)}[-16 \pi g t_{LL}^{i j}]
	  &=& \frac{1}{4}
	  \left(\mbox{}_{(4)} \bar{h}^{\tau \tau ,i}
	        \mbox{}_{(4)} \bar{h}^{\tau \tau ,j}
	  - \frac{1}{2} \delta^{ij}
	        \mbox{}_{(4)} \bar{h}^{\tau \tau ,k}
	        \mbox{}_{(4)} \bar{h}^{\tau \tau}\mbox{}_{,k}\right).
\label{tLL4ij}
    \end{eqnarray}
$O(\epsilon^6)$ 
	  \begin{eqnarray}
       \mbox{}_{(6)}[-16 \pi g t_{LL}^{\tau \tau}]
	   &=& -\frac{7}{8} \mbox{}_{(4)} \bar{h}^{\tau \tau ,k}
	        \mbox{}_{(4)} \bar{h}^{\tau \tau}\mbox{}_{,k},
\label{tLL6tt}	 \\
 	    \mbox{}_{(6)}[-16 \pi g t_{LL}^{\tau i}]
	   &=& 2 \mbox{}_{(4)}\bar{h}^{\tau \tau}\mbox{}_{,k}
	       \mbox{}_{(4)}\bar{h}^
		   {\tau {\scriptscriptstyle [}k ,i{\scriptscriptstyle ]}}
		   - \frac{3}{4}\mbox{}_{(4)}\bar{h}^{\tau k}\mbox{}_{,k}
		                \mbox{}_{(4)}\bar{h}^{\tau \tau,i},
\label{tLL6ti}		\\	
	  \mbox{}_{(6)}[-16 \pi g t_{LL}^{i j}]
	  &=& 4\mbox{}_{(4)}\bar{h}^
	  {\tau {\scriptscriptstyle [}i ,k{\scriptscriptstyle ]}}
	    \mbox{}_{(4)}\bar{h}^{\tau}\mbox{}_{{\scriptscriptstyle [ }k}
		\mbox{}^{,j{\scriptscriptstyle ]}}
		- 2\mbox{}_{(4)}\bar{h}^{\tau \tau ,{\scriptscriptstyle (}i}
		   \mbox{}_{(4)}\bar{h}^{j{\scriptscriptstyle )}k}\mbox{}_{,k}
	\nonumber \\ 
  	   \mbox{} &+& \frac{1}{2}
		\mbox{}_{(4)}\bar{h}^{\tau \tau ,{\scriptscriptstyle (}i}
	       \mbox{}_{(6)}\bar{h}^
		   {|\tau \tau | ,j{\scriptscriptstyle )}}
		+   \frac{1}{2}
		\mbox{}_{(4)}\bar{h}^{\tau \tau ,{\scriptscriptstyle (}i}
	       \mbox{}_{(4)}\bar{h}^
		   {|k}\mbox{}_{k|}\mbox{}^{,j{\scriptscriptstyle )}}
		-  \frac{1}{2}\mbox{}_{(4)}\bar{h}^{\tau \tau}
		\mbox{}_{(4)}\bar{h}^{\tau \tau ,i}
		\mbox{}_{(4)}\bar{h}^{\tau \tau ,j}   
     \nonumber \\
	   \mbox{} &+& \delta^{ij}
		\left[
					-\frac{3}{8}\mbox{}_{(4)}\bar{h}^{\tau k}\mbox{}_{,k}
					            \mbox{}_{(4)}\bar{h}^{\tau l}\mbox{}_{,l}
                    + \mbox{}_{(4)}\bar{h}^{\tau \tau}\mbox{}_{,k}
					  \mbox{}_{(4)}\bar{h}^{k l}\mbox{}_{,l}
                    + \mbox{}_{(4)}\bar{h}^{\tau k, l}
					  \mbox{}_{(4)}\bar{h}^{\tau}
					  \mbox{}_{{\scriptscriptstyle [}k,
					  l{\scriptscriptstyle ]}} \right.
			\nonumber		  \\ 
        \mbox{} &-& \left. 
					 \frac{1}{4}\mbox{}_{(4)}\bar{h}^{\tau \tau}\mbox{}_{,k}
					            \mbox{}_{(6)}\bar{h}^{\tau \tau,k}
                    -\frac{1}{4}\mbox{}_{(4)}\bar{h}^{\tau \tau,k}
					  \mbox{}_{(4)}\bar{h}^{l}\mbox{}_{l,k}
                    + \frac{1}{4}
					  \mbox{}_{(4)}\bar{h}^{\tau \tau}
					  \mbox{}_{(4)}\bar{h}^{\tau\tau}\mbox{}_{,k}
					  \mbox{}_{(4)}\bar{h}^{\tau\tau,k}
		            \right].
\label{tLL6ij}		 
	  \end{eqnarray}  
$O(\epsilon^7)$ 
      \begin{eqnarray}
	   \mbox{}_{(7)}[-16 \pi g t_{LL}^{\tau \tau}] &=& 0,
	   \nonumber \\
	   \mbox{}_{(7)}[-16 \pi g t_{LL}^{\tau i}] &=& 0,
	   \nonumber \\
	   \mbox{}_{(7)}[-16 \pi g t_{LL}^{i j}] &=&
	   \frac{1}{2}\left[
	   \mbox{}_{(4)}\bar{h}^{\tau\tau,{\scriptscriptstyle (}i}
	   \mbox{}_{(5)}\bar{h}^{|k}\mbox{}_{k|}\mbox{}^{,j{\scriptscriptstyle )}}
      +\mbox{}_{(4)}\bar{h}^{\tau\tau,{\scriptscriptstyle (}i}
	   \mbox{}_{(7)}\bar{h}^{|\tau\tau|,j{\scriptscriptstyle )}} 
    \right. \nonumber \\
    \mbox{} &-& \left.
    \frac{1}{2}\delta^{ij}\left\{
       \mbox{}_{(4)}\bar{h}^{\tau\tau,l}
	   \mbox{}_{(5)}\bar{h}^{k}\mbox{}_{k,l}
      +\mbox{}_{(4)}\bar{h}^{\tau\tau,k}
	   \mbox{}_{(7)}\bar{h}^{\tau\tau}\mbox{}_{,k}\right\}
	  \right].
      \end{eqnarray}
$O(\epsilon^8)$ 
 \begin{eqnarray}
	   \mbox{}_{(8)}[-16 \pi g t_{LL}^{\tau \tau}] &=& 
       -\frac{7}{4}\mbox{}_{(4)}\bar{h}^{\tau\tau,k}
	   \mbox{}_{(6)}\bar{h}^{\tau\tau}\mbox{}_{,k}
	   \nonumber \\
	   \mbox{} &-&
		\frac{3}{8}\mbox{}_{(4)}\bar{h}^{\tau k}\mbox{}_{,k}
	   \mbox{}_{(4)}\bar{h}^{\tau l}\mbox{}_{,l}
	   +\mbox{}_{(4)}\bar{h}^{\tau k,l}
	   \mbox{}_{(4)}\bar{h}^{\tau}\mbox{}_{{\scriptscriptstyle
	   (}k,l{\scriptscriptstyle )}}
	   \nonumber \\
       \mbox{} &+&
       \frac{1}{4}\mbox{}_{(4)}\bar{h}^{\tau\tau,k}
	   \mbox{}_{(4)}\bar{h}^{l}\mbox{}_{l,k}
	  +\mbox{}_{(4)}\bar{h}^{\tau\tau}\mbox{}_{,k}
	   \mbox{}_{(4)}\bar{h}^{k l}\mbox{}_{,l}
	   \nonumber \\
      \mbox{} &+&
      \frac{7}{8}\mbox{}_{(4)}\bar{h}^{\tau\tau}
	  \mbox{}_{(4)}\bar{h}^{\tau\tau,k}\mbox{}_{(4)}\bar{h}^{\tau\tau}
	  \mbox{}_{,k}, 
\label{tLL8tt}\\
     \mbox{}_{(8)}[-16 \pi g t_{LL}^{\tau i}] &=& 
      2 \mbox{}_{(4)}\bar{h}^{\tau}\mbox{}_{k,l}
	    \mbox{}_{(4)}\bar{h}^{k{\scriptscriptstyle[}i,l{\scriptscriptstyle
		]}}
	+ \mbox{}_{(4)}\bar{h}^{\tau i}\mbox{}_{,k}
	   \mbox{}_{(4)}\bar{h}^{kl}\mbox{}_{,l}
	+ \frac{1}{4}\mbox{}_{(4)}\bar{h}^{\tau l}\mbox{}_{,l}
	   \mbox{}_{(4)}\bar{h}^{k}\mbox{}_{k}\mbox{}^{i}
\nonumber \\
    \mbox{} &-& \frac{1}{4}\mbox{}_{(4)}\bar{h}^{\tau\tau,i}
	 \mbox{}_{(4)}\bar{h}^{l}\mbox{}_{l,\tau}
\nonumber \\
     \mbox{} &+&
	  2
	  \mbox{}_{(6)}\bar{h}^
	  {\tau {\scriptscriptstyle[}k,i{\scriptscriptstyle ]}}
	  \mbox{}_{(4)}\bar{h}^{\tau \tau}\mbox{}_{,k}
	  -\frac{3}{4}\mbox{}_{(6)}\bar{h}^{\tau k}\mbox{}_{,k}
	  \mbox{}_{(4)}\bar{h}^{\tau \tau,i}
\nonumber \\
     \mbox{} &+&
	  2 \mbox{}_{(6)}\bar{h}^{\tau \tau}\mbox{}_{,k}
	  \mbox{}_{(4)}\bar{h}^
	  {\tau {\scriptscriptstyle [}k,i{\scriptscriptstyle ]}}
      -\frac{3}{4}\mbox{}_{(6)}\bar{h}^{\tau \tau,i}
	  \mbox{}_{(4)}\bar{h}^{\tau k}\mbox{}_{,k}
\nonumber \\
  \mbox{} &+&
   2\mbox{}_{(4)}\bar{h}^{\tau\tau}
   \mbox{}_{(4)}\bar{h}^{\tau\tau}
   \mbox{}_{,l}\mbox{}_{(4)}\bar{h}^
   {\tau{\scriptscriptstyle[}i,l{\scriptscriptstyle ]}}
 + \frac{3}{4}\mbox{}_{(4)}\bar{h}^{\tau\tau}
   \mbox{}_{(4)}\bar{h}^{\tau\tau, i}
   \mbox{}_{(4)}\bar{h}^
   {\tau l}\mbox{}_{l}
\nonumber \\
    \mbox{} &-& \frac{1}{4}
	 \mbox{}_{(4)}\bar{h}^{\tau k}
	 \mbox{}_{(4)}\bar{h}
	 ^{\tau\tau}\mbox{}_{,k}\mbox{}_{(4)}\bar{h}^{\tau\tau,i}
	 + \frac{1}{8}\mbox{}_{(4)}\bar{h}^{\tau i}
	 \mbox{}_{(4)}\bar{h}
	 ^{\tau\tau}\mbox{}_{,l}\mbox{}_{(4)}\bar{h}^{\tau\tau,l},
\label{tLL8ti}\\
\mbox{}_{(8)}[-16 \pi g t_{LL}^{i j}]
&=&
- 2\mbox{}_{(4)}\bar{h}^{\tau\tau,{\scriptscriptstyle (}i}
 \mbox{}_{(6)}\bar{h}^{j{\scriptscriptstyle )}l}\mbox{}_{,l}
+\frac{1}{2}\mbox{}_{(4)}\bar{h}^{\tau\tau,{\scriptscriptstyle (}i}
 \mbox{}_{(6)}\bar{h}^{|l}\mbox{}_{l|}\mbox{}^{,j{\scriptscriptstyle )}} 
\nonumber \\
\mbox{} &+&
 2\mbox{}_{(6)}\bar{h}^{\tau l,{\scriptscriptstyle (}i}
  \mbox{}_{(4)}\bar{h}^{j{\scriptscriptstyle )}\tau}\mbox{}_{,l}
-2\mbox{}_{(6)}\bar{h}^{\tau {\scriptscriptstyle (}i,|l|}
  \mbox{}_{(4)}\bar{h}^{j{\scriptscriptstyle )}\tau}\mbox{}_{,l}
+ 2\mbox{}_{(4)}\bar{h}^{\tau l,{\scriptscriptstyle (}i}
  \mbox{}_{(6)}\bar{h}^{j{\scriptscriptstyle )}\tau}\mbox{}_{,l}
-2\mbox{}_{(6)}\bar{h}^{\tau}\mbox{}_{l}\mbox{}^{,{\scriptscriptstyle (}i}
  \mbox{}_{(4)}\bar{h}^{|\tau l|,j{\scriptscriptstyle )}}
\nonumber \\
\mbox{} &+&
 \frac{1}{2}\mbox{}_{(4)}\bar{h}^{\tau\tau,{\scriptscriptstyle (}i}
            \mbox{}_{(8)}\bar{h}^{|\tau\tau|,j{\scriptscriptstyle )}}
\nonumber \\
\mbox{} &+&  \delta^{i j}
 \left[- \frac{3}{4}
		 \mbox{}_{(4)}\bar{h}^{\tau k}\mbox{}_{,k}
         \mbox{}_{(6)}\bar{h}^{\tau l}\mbox{}_{,l}
        +2\mbox{}_{(4)}\bar{h}^{\tau k,l}
		  \mbox{}_{(6)}\bar{h}^{\tau}
		  \mbox{}_{{\scriptscriptstyle [}k,l{\scriptscriptstyle ]}}
\right. \nonumber \\
\mbox{} &+& \left.
		 \mbox{}_{(4)}\bar{h}^{\tau \tau}\mbox{}_{,l}
         \mbox{}_{(6)}\bar{h}^{l k}\mbox{}_{,k}
		-\frac{1}{4}
		 \mbox{}_{(4)}\bar{h}^{\tau \tau,k} 
         \mbox{}_{(6)}\bar{h}^{l}\mbox{}_{l,k}
\right. \nonumber \\
\mbox{} &-& \left. 
         \frac{1}{4}\mbox{}_{(4)}\bar{h}^{\tau \tau,k}
		 \mbox{}_{(8)}\bar{h}^{\tau \tau}\mbox{}_{,k}
\right]
\nonumber \\
\mbox{}  &+& ({\rm Terms\, independent\, of\, the\, spin\, and\,
 quadrupole\, moment}).
\label{tLL8ij}
 \end{eqnarray}

Next, we show here the field $\bar h^{\mu\nu}$ expanded in $\epsilon$.
The derivations can be seen in \cite{Futamase87}, \cite{JS}, \cite{BFP}
and partly in the section V and appendix A.
We do not show the field of order
$\epsilon^5$ and $\epsilon^7$, since these depend only on time
\cite{Futamase87} and hence have no effect on the equation of motion
up to the order relevant to this paper \cite{InvalidExpression}.\\

\begin{eqnarray}
\bar{h}^{\tau \tau} &=& \epsilon^4 \mbox{}_{(4)}\bar{h}^{\tau\tau} 
                       +\epsilon^6 \mbox{}_{(6)}\bar{h}^{\tau\tau}
                       +\epsilon^7 \mbox{}_{(7)}\bar{h}^{\tau\tau} 
                       +\epsilon^8 \mbox{}_{(8)}\bar{h}^{\tau\tau}
 					   + O(\epsilon^9), 
\\
\mbox{}_{(4)}\bar{h}^{\tau\tau} &=&
 4 \sum_{A=1,2} \frac{\bar{P}^{\tau}_A}{r_A},
\label{InvalidExpressionB}
\\ 
\mbox{}_{(6)}\bar{h}^{\tau\tau} &=&
              -2 \sum_{A=1,2}\frac{\bar{m}_A}{r_A}
			     \{(\vec{v}_A\cdot\vec{n}_A)^2-v_A^2\}
                   +2\frac{\bar{m}_1 \bar{m}_2}{r_{12}^2}
				   \vec{n}_{12}\cdot(\vec{n}_1-\vec{n}_2)
\nonumber \\
\mbox{} 
&+&  7 \sum_{A=1,2}\frac{\bar{m}_A^2}{r_A^2}
+ 14\frac{\bar{m}_1 \bar{m}_2}{r_1 r_2}
           -14\frac{\bar{m}_1 \bar{m}_2}{r_{12}} \sum_{A=1,2}\frac{1}{r_A},
\\ 
\mbox{}_{(8)}\bar{h}^{\tau\tau} &=&
6\sum_{A=1,2} \frac{r_A^k r_A^l}{r_A^5}{\hat I}_A^{kl} 
+4 \sum_{A=1,2} \frac{r_A^k}{r_A^3} M_A^{kl} v_A^l \nonumber \\
\mbox{} &+& ({\rm Terms\, independent\, of\, the\, spin\, and\,
 quadrupole\, moment}).
\end{eqnarray}

\begin{eqnarray}
\bar{h}^{\tau i} &=& \epsilon^4 \mbox{}_{(4)}\bar{h}^{\tau i} +
                        \epsilon^6 \mbox{}_{(6)}\bar{h}^{\tau i}
						+ O(\epsilon^7), 
\\
\mbox{}_{(4)}\bar{h}^{\tau i} &=&
 4 \sum_{A=1,2} \frac{\bar{P}^{\tau}_A v^i_A}{r_A},
\\ 
\mbox{}_{(6)}\bar{h}^{\tau i} &=&
              2 \sum_{A=1,2}\frac{r_A^k}{r_A^3}M_A^{k i}
\nonumber \\
\mbox{} &+& ({\rm Terms\, independent\, of\, the\, spin\, and\,
 quadrupole\, moment}).
\label{hti6BandNC}
\end{eqnarray}

\begin{eqnarray}
\bar{h}^{i j} &=& \epsilon^4 \mbox{}_{(4)}\bar{h}^{i j}
                       +\epsilon^5 \mbox{}_{(5)}\bar{h}^{i j}
                       +\epsilon^6 \mbox{}_{(6)}\bar{h}^{i j}
					   + O(\epsilon^7),
\\
\mbox{}_{(4)}\bar{h}^{i j} &=& 
4 \sum_{A=1,2} \frac{\bar{P}^{\tau}_A v_A^i v_A^j}{r_A}
\nonumber \\
\mbox{} &+& 
 \sum_{A=1,2}\frac{\bar{m}_A^2}{r_A^4}r_A^i r_A^j
-\frac{8 \bar{m}_1 \bar{m}_2}{r_{12}S}n_{12}^i n_{12}^j
\nonumber \\
\mbox{} &-&
8\left[\delta^{i}\mbox{}_{k}\delta^{j}\mbox{}_{l}
  - \frac{1}{2}\delta^{i j}\delta_{k l}\right]
\frac{\bar{m}_1 \bar{m}_2}{S^2}
(\vec{n}_{12}-\vec{n}_{1})^{{\scriptscriptstyle (}k}
                   (\vec{n}_{12}+\vec{n}_{2})^{l{\scriptscriptstyle )}},  
\\
 \mbox{}_{(6)}\bar{h}^{ij} &=&
4 \sum_{A=1,2} \frac{r_A^k}{r_A^3} M_A^{k {\scriptscriptstyle (}i}
v_A^{j {\scriptscriptstyle )}} \nonumber \\
\mbox{} &+& ({\rm Terms\, independent\, of\, the\, spin\, and\,
 quadrupole\, moment}).
\end{eqnarray}

\newpage

\begin{table}[h]
 
\begin{center}

\begin{tabular}{c c c} \hline
 Near zone & Near zone  & Body zone  \\ 
 $(t,x^i)$ & $(\tau,x^i)$ & $(\tau,\alpha^{\underline{i}})$ \\
\hline 
$\mbox{}$ & & \\  
$T_A^{tt} \sim \epsilon^{-4} $  & $T_A^{\tau\tau} \sim \epsilon^{-2}$
 & $T_A^{\tau\tau} \sim \epsilon^{-2}$  \\
$T_A^{ti} \sim \epsilon^{-3} $  & $T_A^{\tau i} \sim \epsilon^{-2}$
 & $T_A^{\tau \underline{i}} \sim \epsilon^{-4}$  \\
$T_A^{ij} \sim \epsilon^{-4} $  & $T_A^{ij} \sim \epsilon^{-4}$
 & $T_A^{\underline{i}\underline{j}} \sim \epsilon^{-8}$ \\    
 \hline
\end{tabular}

\end{center}
 
\caption[scalings]{Scalings of the stress energy tensor
 $T_A^{\mu\nu}$ of matters on the initial
 hypersurface in various coordinates.} 

\label{TableOfScalings}

\end{table}

\begin{table}[h]
 
\begin{center}

\begin{tabular}{c|c c c} \hline
& Newtonian order & 1 PN order  & 2 PN order  \\ 
\hline 
$\mbox{}$ &  &  & \\
$\bar{h}^{\tau\tau}$ & $\epsilon^4$ & $\epsilon^6$ & $\epsilon^8$ \\
$\bar{h}^{\tau i}$ & $-$ & $\epsilon^4$ & $\epsilon^6$ \\
$\bar{h}^{ij}$ & $-$ & $\epsilon^4$ & $\epsilon^6$ \\
 \hline
\end{tabular}

\end{center}
 
\caption[scalings]{Required order of the gravitational 
field to obtain an equation of motion (EOM) up to a certain order.
For example, to obtain the Newtonian EOM, we have to calculate
 $\bar{h}^{\tau\tau}$ up to $\epsilon^4$.} 

\label{TableOfGF}

\end{table}

\begin{table}[h]
 
\begin{center}

\begin{tabular}{c|c c c}
\hline
  & Newtonian order & 1 PN order  & 2 PN order  \\ 
\hline
$\mbox{}$ & & & \\
$t_{LL}^{\tau\tau}$ & $-$ & $\epsilon^6$ & $\epsilon^8$ \\
$t_{LL}^{\tau i}$ & $-$ & $\epsilon^6$ & $\epsilon^8$ \\
$t_{LL}^{ij}$ & $\epsilon^4$ & $\epsilon^6$ & $\epsilon^8$
 \\ \hline
\end{tabular}

\end{center}
 
\caption[scalings]{Required order of the Landau-Lifshitz pseudotensor
 to obtain an equation of motion (EOM) up to a certain order.
For example, to obtain the Newtonian EOM, we have to calculate
 $t^{ij}_{LL}$ up to $\epsilon^4$.} 

\label{TableOftLL}

\end{table}


\begin{thebibliography}{999}
\bibitem{Paczynski}B. Paczynski, Astrophys. J. Lett. {\bf 43}, 308 (1986). 
\bibitem{Abramovici}A. Abramovici et al., Science, {\bf 256}, 325 (1992). 
\bibitem{Bradaschia}C. Bradaschia et al., Nucl. Instrum. Method Phys. 
Res. Sect. {\bf A289}, 518 (1990). 
\bibitem{Kuroda}K. Kuroda et. al., in {\it Proceedings of the 
international conference on gravitational waves: Sources and Detections}, 
edited by I. Ciufolini and F. Fidecard 
(World Scientific, Singapore, 1997), p. 100. 
\bibitem{Hough}J. Hough, in {\it Proceedings of the Sixth Marcel 
Grossmann Meeting}, edited by H. Sato and T. Nakamura
(World Scientific, Singapore, 1992), p.192.
\bibitem{BDIWW}L. Blanchet, T. Damour, B. R. Iyer, C. M. Will 
and A. G. Wiseman, Phys. Rev. Lett. {\bf 74}, 3515 (1995).
\bibitem{Blanchet}L. Blanchet, in {\it Proceedings of Les Houches School: 
Relativistic gravitation and gravitational radiation}, edited by 
J. A. Marck and J. P. Lasota (Cambridge Univ. Press, Cambridge, 1996)
\bibitem{MSSTT}Y. Mino, M. Sasaki, M. Shibata, H. Tagoshi and T. Tanaka, 
Prog. Theor. Phys. Suppl. {\bf 128}, 1 (1997). 
\bibitem{AF}H. Asada and T. Futamase, Prog. Theor. Phys. Suppl. 
{\bf 128}, 123 (1997).
\bibitem{CE}S. Chandrasekhar and F. P. Esposito, Astrophys. J. 
{\bf 160}, 153 (1970). 
\bibitem{DD81a}T. Damour and N. Deruelle, Phys. Lett. {\bf 87A}, 81 (1981).
\bibitem{DD81b}T. Damour and N. Deruelle, C. R. Acad. Sci. Paris 
{\bf 293} II, 537 (1981); {\bf 293} II, 877 (1981).
\bibitem{Damour}T. Damour, C. R. Acad. Sci. Paris {\bf 294} II, 1355 (1982). 
\bibitem{Kopejkin}S. M. Kopejkin, Sov. Astron. {\bf 29}, 516 (1985). 
\bibitem{BFP}L. Blanchet, G. Faye and B. Ponsot, 
Phys. Rev. {\bf D58}, 124002 (1998). 
\bibitem{FS83}T. Futamase and B. F. Schutz, Phys. Rev. 
{\bf D28}, 2363 (1983). 
\bibitem{JS}R. Jaranowski and G. Schafer, Phys. Rev. 
{\bf D57}, 7274 (1998); {\bf D57}, 5948 (1998). 
\bibitem{GII}A. Gopakumar, B. R. Iyer and S. Iyer, 
Phys. Rev. {\bf D55}, 6030 (1997).
\bibitem{BDDIM}L. Bel, T. Damour, N. Deruelle, J. Ibanez and J. Martin, 
Gen. Rel. Grav. {\bf 13}, 963 (1981).
\bibitem{GK}L. P. Grischuk and S. M. Kopejkin, Sov. Astron. Lett. 
{\bf 9}, 230 (1983). 
\bibitem{DEath}P. D. D'Eath, Phys. Rev. {\bf D11}, 1387 (1975a); 
{\bf D11}, 2183 (1975b). 
\bibitem{TH}K. S. Thorne and J. B. Hartle, Phys. Rev. {\bf D31}, 1815 (1985).
 \bibitem{Futamase85}T. Futamase, Phys. Rev. {\bf D32}, 2566 (1985).
\bibitem{Futamase87}T. Futamase, Phys. Rev. {\bf D36}, 321 (1987).
\bibitem{TN94} H. Tagoshi and T. Nakamura, Phys. Rev. {\bf D49}, 4016 (1994). 
 \bibitem{Schutz80}B. F. Schutz, Phys. Rev. {\bf D22}, 249 (1980).
\bibitem{Futamase83}T. Futamase, Phys. Rev. {\bf D28}, 2373 (1983).
\bibitem{Kidder}L. E. Kidder, Phys. Rev. {\bf D52}, 821 (1995).
\bibitem{BK}V. A. Brumberg and S. M. Kopejkin, Nuovo Cimento B {\bf 103}
 63 (1989).
\bibitem{DSX}T. Damour, M. Soffel and C. Xu, Phys. Rev.
{\bf D43}, 3273 (1991);{\bf D45}, 1017 (1992);{\bf D47}, 3124 (1993);
 {\bf D49}, 618 (1994) 
 \bibitem{LL}L. D. Landau and E. M. Lifshitz, {\it The Classical Theory 
of Fields} (Oxford: Pergamon 1962). 
 \bibitem{InvalidExpression} Mathematically,  equations
 (\ref{InvalidExpressionA}) and (\ref{InvalidExpressionB}) are
 wrong about the ordering. But it was easier to calculate with these
 expressions than those in a correct ordering, so we adopted the former
ones.
\end{thebibliography}
\end{document}